\documentclass[12pt]{article}
\usepackage{epsfig}

\def\beq{\begin{equation}}
\def\eeq#1{\label{#1}\end{equation}}
\def\eeqn{\end{equation}}

\def\beqa{\begin{eqnarray}}
\def\eeqa#1{\label{#1}\end{eqnarray}}
\def\eeqan{\end{eqnarray}}

\let\bar=\overbar

\def\Dslash{\not{\hbox{\kern-4pt $D$}}}
\def\dslash{\not{\hbox{\kern-2pt $\del$}}}

\def\msb{{\bar{\ssstyle M \kern -1pt S}}}

\newcommand{\gou}{\raisebox{-0.0\totalheight}{\includegraphics[scale=.4]{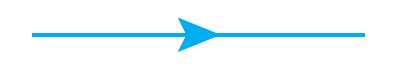}}}
\newcommand{\god}{\raisebox{-0.0\totalheight}{\includegraphics[scale=.4]{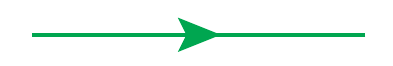}}}
\newcommand{\gol}{\raisebox{-0.0\totalheight}{\includegraphics[scale=.4]{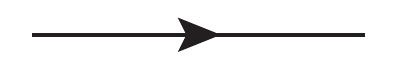}}}
\newcommand{\goi}{\raisebox{-0.3\totalheight}{\includegraphics[scale=.4]{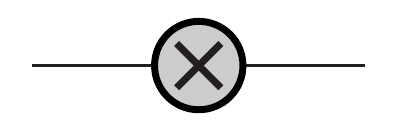}}}

\newcommand{\gdsi}{\raisebox{-0.4\totalheight}{\includegraphics[scale=.3]{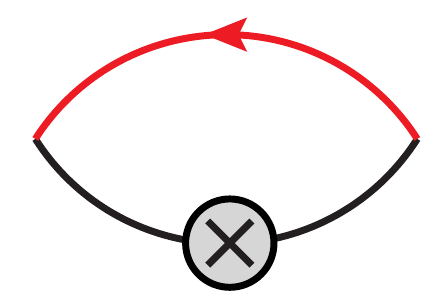}}}
\newcommand{\gdsu}{\raisebox{-0.4\totalheight}{\includegraphics[scale=.3]{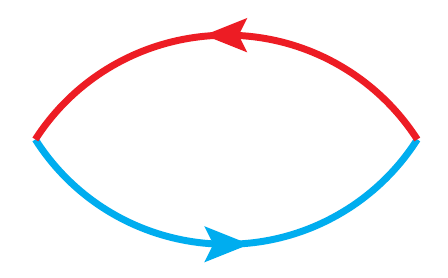}}}
\newcommand{\gdsd}{\raisebox{-0.4\totalheight}{\includegraphics[scale=.3]{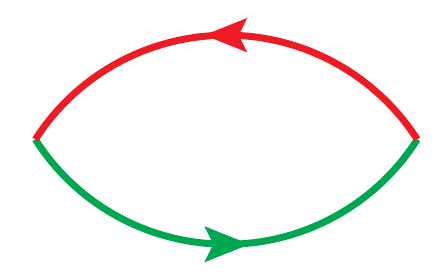}}}
\newcommand{\gdsl}{\raisebox{-0.4\totalheight}{\includegraphics[scale=.3]{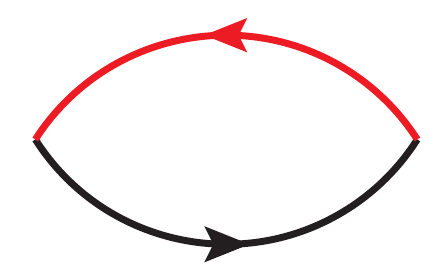}}}

\def\Title#1{\begin{center} {\Large {\bf #1} } \end{center}}

\begin{document}

\Title{Lattice calculation of isospin corrections to $K\ell2$ and $K\ell3$ decays}

\bigskip\bigskip


\begin{raggedright}  

{\it Nazario Tantalo\index{Tantalo, N.}\\
Universit\`a di Roma ``Tor Vergata'',\\
INFN sezione di Roma ``Tor Vergata'',\\
Via della Ricerca Scientifica 1, I-00133 Roma, ITALY}\\
\end{raggedright}

\bigskip\bigskip

\begin{center}
{\small\bf Abstract}
\end{center}
{\small
In this talk I discuss the theoretical issues associated with lattice calculations of isospin breaking corrections to hadronic matrix elements. I concentrate on the calculation of QCD isospin breaking effects for the $K\ell2$ and $K\ell3$ decay rates and illustrate the recent lattice results obtained by the RM123 collaboration.
}

\bigskip\bigskip
{\it\small
\noindent Proceedings of CKM 2012, the 7th International Workshop on the CKM Unitarity Triangle, University of Cincinnati, USA, 28 September - 2 October 2012 
}

\section{Introduction}
Isospin is an approximate symmetry of fundamental interactions. The two lightest quarks, the up and the down, have different masses and different electric charges but it happens that their mass difference is a small parameter with respect to the confinement scale $\Lambda_{QCD}$ and that electromagnetic isospin breaking effects are also small on hadronic observables because at low energy $\alpha_{em}$ is much smaller than $\alpha_s$. For these reasons most of the theoretical predictions on phenomenologically relevant hadronic observables have been derived by assuming the exact validity of isospin symmetry. This is also the case for most of the non-perturbative theoretical predictions on hadronic matrix elements obtained over the years by performing lattice QCD simulations.

In this talk I will concentrate on the lattice calculation of isospin breaking corrections to the hadronic matrix elements entering the leptonic ($K\ell2$) and semileptonic ($K\ell3$) decay rates of the kaons. These are among the golden--plated quantities for lattice QCD calculations: these involve a single hadron into initial and/or final states. Furthermore, two and three--point functions of pseudoscalar mesons interpolating operators can be obtained with high statistical accuracy by using a reasonably small number of gauge configurations. 

In order to highlight the importance of a non--perturbative first principle calculation of the isospin breaking effects on these quantities I quote below the FLAG~\cite{Colangelo:2010et} averages of the lattice results for the ratio of the kaons ($F_K$) and pions ($F_\pi$) decay constants and for the form factor parametrizing the semileptonic decay of a kaon into a pion ($f_+^{K\pi}(q^2)$),
\begin{eqnarray}
\left. \frac{F_K}{F_\pi} \right\vert^{FLAG}=1.193(5)\; ,
\qquad
\left. f_+^{K\pi}(0) \right\vert^{FLAG}=0.956(8) \; .
\label{eq:fflatt}
\end{eqnarray}
These results assume exact isospin symmetry. One may wonder whether the quoted uncertainties ($0.5$\% and $0.8$\% respectively) can be considered reliable in view of all the systematics sources of uncertainties that have to be taken into account when a lattice calculation is performed (unquenching of the heavy flavours, chiral extrapolations, finite volume effects, renormalization, continuum extrapolations, etc.). For a detailed discussion of all the systematics  affecting lattice calculations of $K\ell2$ and $K\ell3$ decay rates see, for example, the contributions of J.~Laiho \index{Laiho, J.} and of A.~J\"uttner \index{Juttner, A.} to these proceedings. Here I argue that the figures quoted above are in fact reliable, by using the argument of the FlaviaNet working group~\cite{Antonelli:2010yf} (see also~\cite{Colangelo:lattice12} and the contribution of V.~Cirigliano \index{Cirigliano, V.} to these proceedings) based on the first--row CKM unitarity test. Indeed, by neglecting $V_{ub}$ because it is smaller than the uncertainties on the other quantities, by using the experimental determinations of the pions and kaons leptonic decay rates together with the experimental determination of the $K\ell3$ decay rate and by using $V_{ud}$ as extracted by the combined experimental and theoretical analysis of $\sim 20$ super--allowed nuclear $\beta$--decays,
\begin{eqnarray}
&& \left\vert \frac{V_{us} F_K}{V_{ud} F_\pi} \right\vert^{exp} = 0.2758(5) \; ,
\qquad \left\vert V_{us} f_+^{K\pi}(0) \right\vert^{exp} = 0.2163(5) \; ,
\nonumber \\
\nonumber \\
&& \left\vert V_{ud} \right\vert^{exp} = 0.97425(22) \; ,
\qquad \left\vert V_{ud} \right\vert^2+\left\vert V_{us} \right\vert^2 = 1 \; ,
\end{eqnarray}
one is able to get a ``measurement'' of the hadronic unknowns,
\begin{eqnarray}
\left. \frac{F_K}{F_\pi} \right\vert^{CKM}=1.1919(57)\; ,
\qquad
\left. f_+^{K\pi}(0) \right\vert^{CKM}=0.9595(46) \; .
\label{eq:ffnolatt}
\end{eqnarray}
The numbers above have been obtained without using lattice inputs and have been corrected for isospin breaking effects by using the effective field theory calculations of refs.~\cite{Kastner:2008ch,Cirigliano:2011tm}. Concerning the QCD isospin breaking effects, i.e. those coming from the difference of the up and down quark masses in absence of electromagnetic interactions, the authors of refs.~\cite{Kastner:2008ch,Cirigliano:2011tm} obtain
\begin{eqnarray}
\left(
\frac{F_{K^+}/F_{\pi^+}}{F_{K}/F_{\pi}}-1
\right)^{QCD,\chi pt}=-0.0022(6) \; ,
\qquad
\left(
\frac{f_+^{K^+\pi^0}(q^2)}{f_+^{K^0\pi^-}(q^2)}-1
\right)^{QCD,\chi pt}=0.029(6) \; .
\label{eq:ffisochi}
\end{eqnarray}

There are some considerations that follow from the comparison of eqs.~(\ref{eq:fflatt}) with eqs.~(\ref{eq:ffnolatt}) and eqs.~(\ref{eq:ffisochi}). The uncertainties quoted by lattice calculations are in fact reliable and, at least for these gold--plated quantities, lattice QCD calculations entered the $1$\% level of accuracy. Furthermore, isospin breaking effects on these quantities are as big as the current lattice errors and cannot be neglected in future lattice calculations, it would be useless otherwise to put efforts in further reducing lattice errors on the isosymmetric quantities. Isospin breaking effects are particularly large in the case of the $K\ell3$ decay rate. By assuming $SU(3)$--flavour symmetry the form factor at vanishing momentum transfer is constrained to be equal to one by the conservation of the vector current and isospin breaking effects are of the same order of magnitude of the deviation of $f_+^{K\pi}(0)$ from this normalization condition. 

In the rest of this talk I describe the method recently developed by the RM123 collaboration~\cite{deDivitiis:2011eh} in order to calculate on the lattice QCD isospin breaking effects and I discuss the results obtained by applying this method. In the last part of the talk I also briefly comment on the lattice calculation of electromagnetic isospin breaking effects.

\section{Leading QCD isospin breaking effects}
The RM123 collaboration has recently proposed~\cite{deDivitiis:2011eh} a method to calculate leading QCD isospin breaking effects on the lattice by starting from gauge configurations generated within the isosymmetric QCD theory ($m_d=m_u$). The method is based on the perturbative expansion of the path-integral with respect to the small parameter $\Delta m_{ud}=(m_d-m_u)/2$,
\begin{eqnarray} 
\langle {\cal O}\rangle 
&\simeq& 
\frac{\int{ D\phi \ {\cal O}\, (1+  \Delta m_{ud} \, \hat S)\, e^{-S_0} }}
{\int{ D\phi \ \, (1+  \Delta m_{ud} \, \hat S) \, e^{-S_0} }}
= 
\frac{\langle {\cal O}\rangle_0 +  \Delta m_{ud} \, \langle {\cal O}\hat S\rangle_0 }
{1+ \Delta m_{ud} \, \langle \hat S\rangle_0 }\, 
\nonumber \\
\nonumber \\
&=& 
\langle {\cal O}\rangle_0 +  \Delta m_{ud} \, \langle {\cal O}\hat S\rangle_0 \, ,
\label{eq:basicofthemethod}
\end{eqnarray}
where $\langle \cdot \rangle_0$ represent the vacuum expectation value in the isospin symmetric theory and $\hat S=\sum_x{[\bar u u - \bar d d](x)}$ is the isospin breaking term. The correction in the denominator vanishes, $\langle \hat S\rangle_0=0$, because of isospin symmetry. In order to take into account first order insertions of $\hat S$ into lattice correlators, one can start by the given observable into the full isospin broken theory, integrate out the quark fields (take the fermionic Wick contractions) and expand the quark propagators at fixed QCD background field with respect to $\Delta m_{ud}$. Diagrammatically this amounts to
\begin{eqnarray}
&&\stackrel{u}{\gou} = \gol + \Delta m_{ud}\goi  +  \cdots \, ,
\nonumber \\
&&\stackrel{d}{\god} = \gol - \Delta m_{ud}\goi  +  \cdots \, ,
\label{eq:props}
\end{eqnarray}
where the propagators with an insertion are simply obtained by calculating the square of the propagator in the isosymmetric theory. By using this simple recipe (see ref.~\cite{deDivitiis:2011eh} for further details and for all the unexplained notation) is easy to show that pion masses and decay constants do not get a first--order QCD isospin breaking correction. Concerning pseudoscalar--pseudoscalar two--point correlators of the kaons, one gets
\begin{eqnarray}
C_{K^+K^-}(t) &=& -\gdsu=   -\gdsl -\Delta m_{ud}\gdsi + {\cal O}(\Delta m_{ud})^2\, ,
\nonumber \\
\nonumber \\
C_{K^0K^0}(t) &=& -\gdsd= -  \gdsl +\Delta m_{ud}\gdsi + {\cal O}(\Delta m_{ud})^2  \, ,   
\label{eq:kpcorr}
\end{eqnarray}
where the strange quark propagator has been drawn in red and where the same color code of eqs.~(\ref{eq:props}) has been used for the up and down propagators. By studying the euclidean time behavior of the ratio of correlators appearing on the right--hand sides of the previous equations,
\begin{eqnarray}
\frac{\Delta C_{KK}(t)}{\Delta m_{ud}} = - \frac{\gdsi}{\gdsl} \; ,
\label{eq:deltakratio}
\end{eqnarray}
it is possible to calculate the leading QCD isospin breaking effects on kaon masses and decay constants.
\begin{figure}[!t]
\begin{center}
\includegraphics[width=0.49\textwidth]{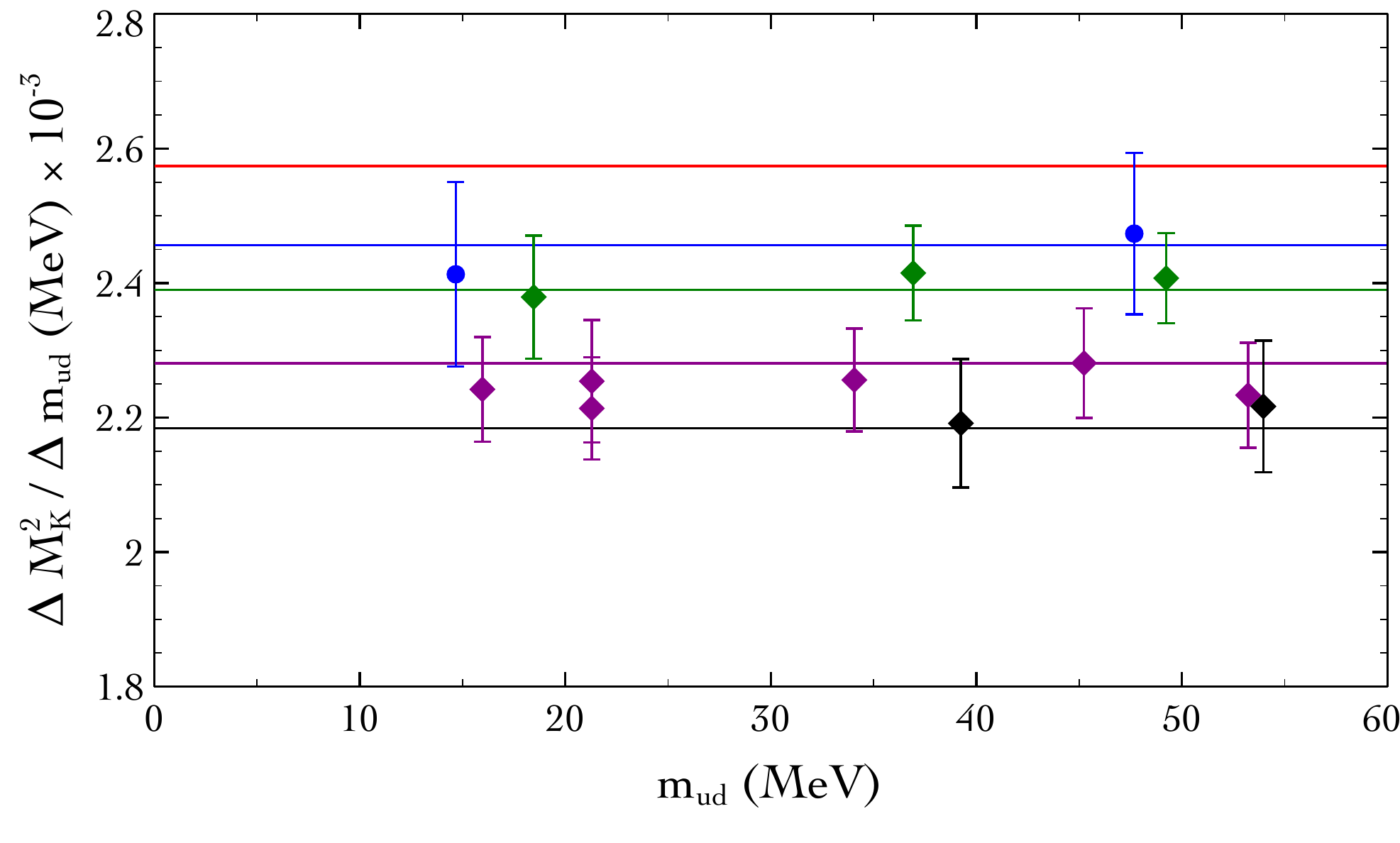}\hfill
\includegraphics[width=0.49\textwidth]{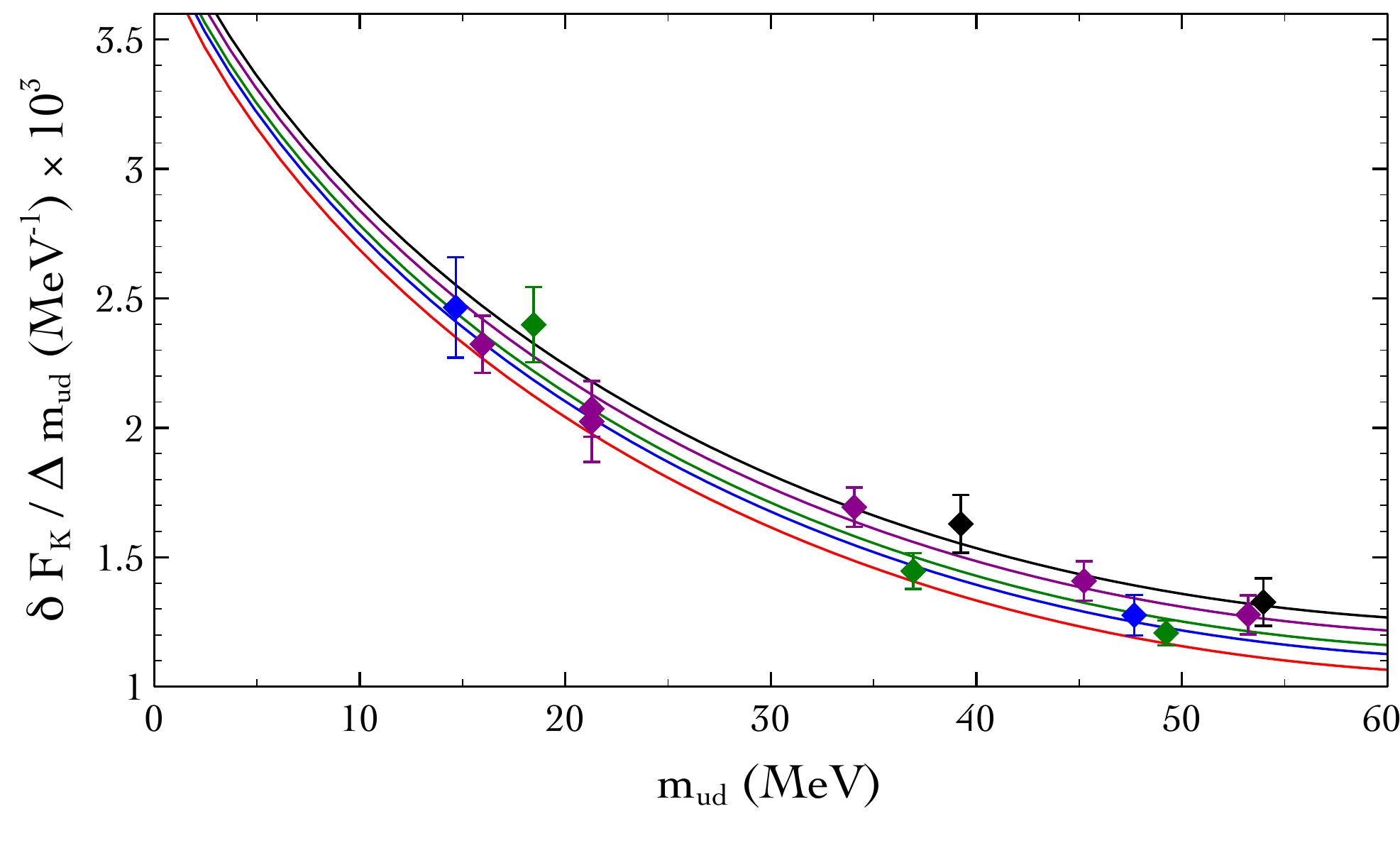}
\caption{\label{fig:Kchiral} \footnotesize
{\it Left panel}: combined chiral and continuum extrapolations of $\Delta M_K^2/\Delta m_{ud}$. 
{\it Right panel}: combined chiral and continuum extrapolations of $\delta F_K/\Delta m_{ud}$. Black points correspond to the coarser lattice spacing, $a=0.098$~fm, dark magenta points correspond to $a=0.085$~fm, green points to $a=0.067$~fm and blue points to $a=0.054$~fm. Red lines are the results of the continuum extrapolations.}
\end{center}
\end{figure}
In Figure~\ref{fig:Kchiral} I show the combined chiral and continuum extrapolations of the lattice data of ref.~\cite{deDivitiis:2011eh} from which the following results
\begin{eqnarray}
\left[\frac{\Delta M_K^2}{\Delta m_{ud}}\right]^{QCD}(\overline{MS},2GeV)
&=&\left[\frac{M_{K^0}^2-M_{K^+}^2}{m_d-m_u}\right]^{QCD} 
=
2.57(8)\times 10^{3}\ \mbox{MeV}\, ,
\nonumber \\
\nonumber \\
\left[\frac{\delta F_K}{\Delta m_{ud}}\right]^{QCD}(\overline{MS},2GeV)
&=& 
2\left[\frac{F_K-F_{K^+}}{F_K(m_d-m_u)}\right]^{QCD}
=3.3(3)\times 10^{-3}\ \mbox{MeV}^{-1},
\label{eq:fitresults}
\end{eqnarray}
are obtained. The results above have been obtained by including in the generation of the gauge configurations the determinants of the two lightest quarks, i.e. in the so--called $n_f=2$ theory. 
By taking the ratio of the previous two results it is possible to express the correction to the $K\ell2$ decay rate, and consequently to $F_K/F_\pi$ (given the fact that $\delta F_\pi=0$ at this order), as a function of the difference of the neutral and charged kaon masses induced by the quark mass difference in QCD. 

It is important to realize that the measured mass difference of the kaons is due to QED as well as to QCD isospin breaking effects and that the first are expected to be not negligible with respect to the latter. Furthermore, the separation of QED from QCD isospin breaking effects is arbitrary and prescription dependent because of the renormalization of the quark masses and of the strong coupling constant required in order to absorb the divergences generated by electromagnetic interactions. QED isospin breaking effects on hadronic quantities can also be calculated on the lattice. The first pioneering calculation has been performed in ref.~\cite{Duncan:1996xy} where lattice QED has been treated in its ``non--compact'' formulation. Later on, several lattice collaborations provided calculations of the electromagnetic mass splittings of the lightest hadrons, see for example refs.~\cite{Basak:2008na,Blum:2010ym,Portelli:2010yn,Ishikawa:2012ix} for recent works on the subject. 

In ref.~\cite{deDivitiis:2013} the RM123 collaboration is going to provide a method to calculate QED+QCD isospin breaking effects by starting from isosymmetric QCD gauge configurations while, for the time being, they have used the following prescription
\begin{eqnarray}
\left[M_{K^0}^2-M_{K^+}^2\right]^{QCD}&=&
\left[M_{K^0}^2-M_{K^+}^2\right]^{exp}-(1+\varepsilon_{\gamma})\left[M_{\pi^0}^2-M_{\pi^+}^2\right]^{exp}
\nonumber \\
\nonumber \\
&=&6.05(63)\times 10^3\ \mbox{MeV}^2\, , \qquad \varepsilon_{\gamma}=0.7(5) \; ,
\end{eqnarray}
with $\varepsilon_{\gamma}$ taken from~\cite{Colangelo:2010et} in order to define the QCD part of the kaons mass difference. The physical meaning of the previous prescription comes from the observation that, according to the well known Dashen's theorem, the mass splittings of the kaons and of the pions have to be equal into the ($m_s=m_d=m_u=0$) chiral limit. The (relatively) small parameter $\varepsilon_{\gamma}$ parametrizes the violation of the Dashen's theorem and can be calculated in a given renormalization scheme by using lattice results. By using this prescription, and the results of eqs.~(\ref{eq:fitresults}), the RM123 collaboration quotes
\begin{eqnarray}
\left[\frac{F_{K^+}/F_{\pi^+}}{F_{K}/F_{\pi}}-1\right]^{QCD} &=& -0.0039(3)(2)
\quad \times \quad 
\frac{\left[M_{K^0}^2-M_{K^+}^2\right]^{QCD}}{6.05\times 10^3\ \mbox{MeV}^2} \, .
\end{eqnarray}
The result above is of the same order of magnitude, though higher, of the chiral perturbation theory estimate of ref.~\cite{Cirigliano:2011tm} (see eqs.~(\ref{eq:ffisochi}) above) and strongly depend upon the value of $\varepsilon_{\gamma}$. It is for this reason that is particularly important to perform combined lattice calculation of QCD+QED isospin breaking effects as it will be done in ref.~\cite{deDivitiis:2013}. 

\begin{figure}[!t]
\begin{center}
\includegraphics[width=0.49\textwidth]{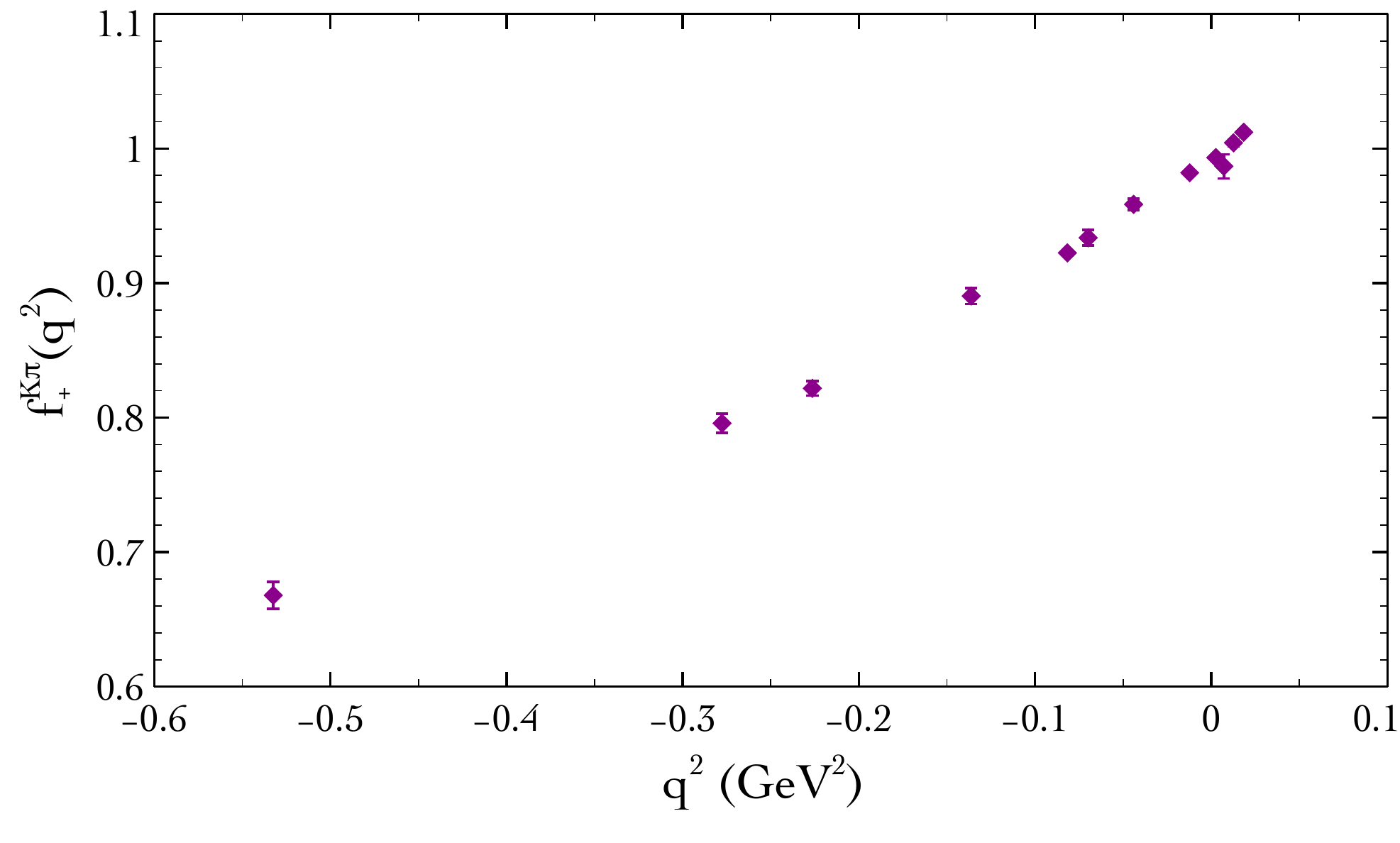}
\includegraphics[width=0.49\textwidth]{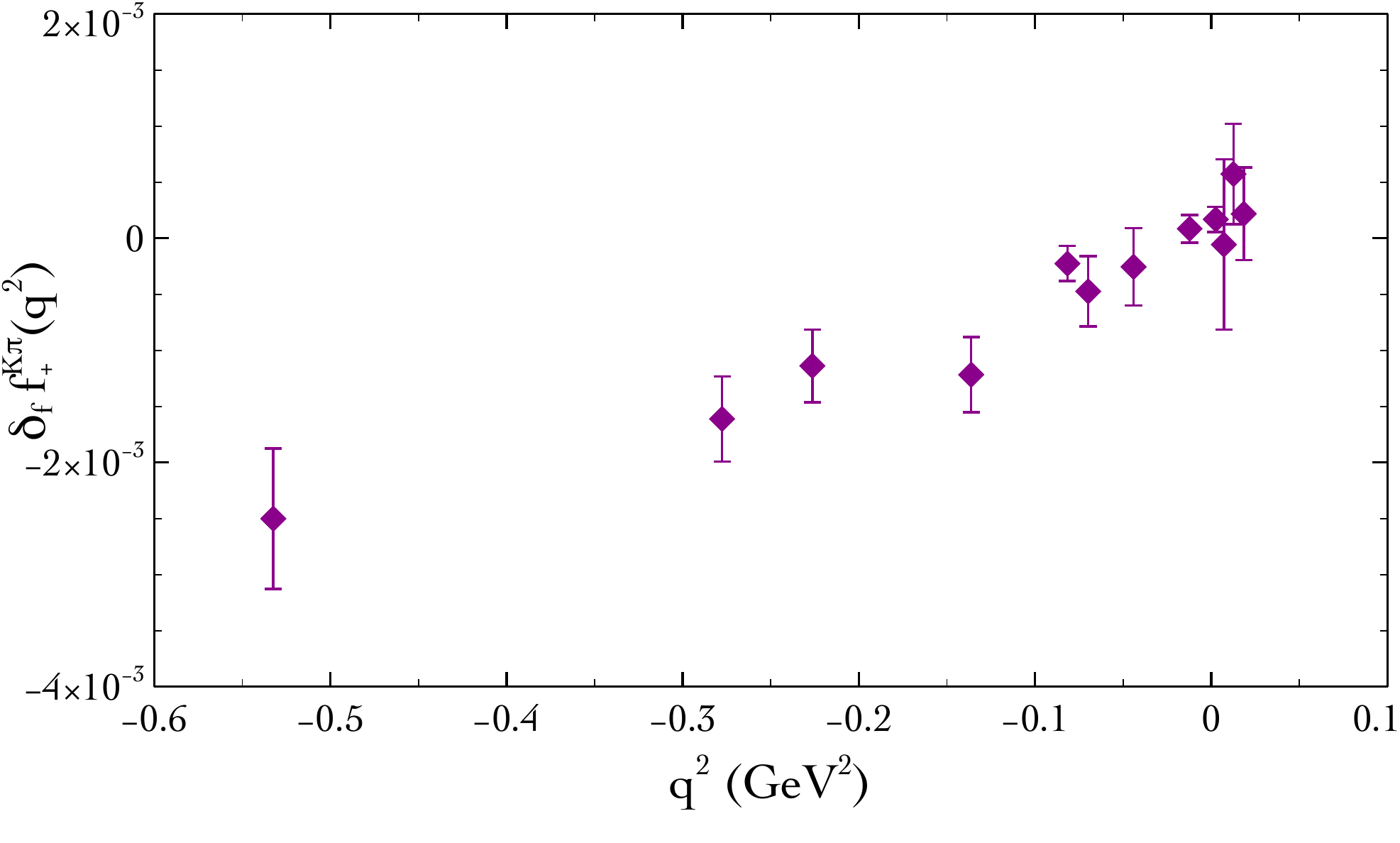}
\caption{\label{fig:deltaf} \footnotesize
{\it Left panel}: results for $f_+^{K\pi}(q^2)$.
{\it Right panel}: results for $\delta_f f_+^{K\pi}(q^2)$. 
The data are obtained at fixed lattice spacing $a=0.085$~fm  and at fixed $am_{ud}^L=0.0064$. 
}
\end{center}
\end{figure}
In the last part of this talk I briefly discuss the issues related to the calculation of QCD isospin breaking corrections to the $K\ell3$ decay rate. By using the same techniques that lead to eq.~(\ref{eq:deltakratio}), in ref.~\cite{deDivitiis:2011eh} it have been derived all the formulae needed in order to perform a lattice calculation of $f_+^{K^+\pi^0}(q^2)/f_+^{K^0\pi^-}(q^2)$. In this case the deviations of the numerator and of the denominator from the common isosymmetric limit $f^{K\pi}(q^2)$ are different. In particular, the formulae for $f_+^{K^+\pi^0}(q^2)$ present quark disconnected (gluon connected) diagrams that are difficult to calculate on the lattice. These disconnected contractions are responsible of the $\pi^0$--$\eta$ mixing and are expected to give a large ``resonant" contribution to the deviation of $f_+^{K^+\pi^0}(q^2)/f_+^{K^0\pi^-}(q^2)$ from one. In ref.~\cite{deDivitiis:2011eh} these disconnected contributions have not been calculated while it has been provided a preliminary estimate of the relative deviation
\begin{eqnarray}
\delta_f f_+^{K\pi}(0)=
\left[ \frac{f_+^{K^0\pi^-}(0)-f_+^{K\pi}(0) }{f_+^{K\pi}(0)}\right]^{QCD}
= 0.85(18)(1)\times 10^{-4} \times  
\frac{\left[M_{K^0}^2-M_{K^+}^2\right]^{QCD}}{6.05\times 10^3\ \mbox{MeV}^2} \, .
\nonumber \\
\label{eq:form}
\end{eqnarray}
In Figure~\ref{fig:deltaf} are shown the corresponding results for the isosymmetric form factor and for the correction as a function of the momentum transfer. The argument concerning the importance of disconnected diagrams is confirmed by the result of eq.~(\ref{eq:form}) that is more than two orders of magnitude smaller of the chiral perturbation theory estimate of the physical correction performed in ref.~\cite{Kastner:2008ch} and quoted into eqs.~(\ref{eq:ffisochi}) above.

\section{Outlooks}
In this talk I have argued that lattice QCD calculations are nowadays sufficiently accurate to be able to provide results for the $K\ell2$ and $K\ell3$ decay rates with sub--percent overall uncertainties. Furthermore, the lattice errors on these quantities can be considered reliable, as emerges from the comparison of world average lattice results with the corresponding quantities obtained from the first--row CKM unitarity test.

At this level of accuracy isospin breaking corrections cannot be neglected anymore because, at least for these quantities, these are of the same order of magnitude of the quoted uncertainties. Isospin breaking corrections can be estimated by recurring to chiral perturbation theory but can also be calculated from first principle lattice calculations. The RM123 collaboration has recently obtained encouraging results for the QCD isospin breaking corrections to the $K\ell2$ decay rate and preliminary results for the $K\ell3$ decay rate.

QED isospin breaking corrections are expected to be as important as QCD ones and are required also for the calculation of the QCD effects in order to cope with the prescription dependent separation of electromagnetic and QCD contributions to the hadrons mass splittings. The QED corrections to the hadron spectrum can also be calculated on the lattice and several lattice collaborations have provided phenomenologically relevant results in the last years.

Concerning QED corrections to weak matrix elements, I want to note in passing that the problem is much more involved with respect to the case of the hadron spectrum. Already in the case of the QED corrections to the $K\ell2$ decay rate, electromagnetic gauge invariance and cancellation of infrared divergences require the inclusion of correlators in which photons connect quark and lepton propagators. For this reason it is not possible to factorize the leptonic tensor in the calculation of the decay rate and more theoretical work is required before we shall be able to provide a first principle lattice calculation of the QED+QCD isospin breaking corrections to $K\ell2$ and $K\ell3$ processes.

\bigskip
\bigskip
I warmly thank my colleagues of the RM123 collaboration for the enjoyable work of the last two years: the lattice results and the theoretical arguments discussed in this talk are the outcome of this work.
I am grateful to the working group conveners and to the organizers of the CKM2012 conference for the opportunity of giving this talk.

\def\Discussion{
\setlength{\parskip}{0.3cm}\setlength{\parindent}{0.0cm}
     \bigskip\bigskip      {\Large {\bf Discussion}} \bigskip}
\def\speaker#1{{\bf #1:}\ }
\def\endDiscussion{}


 
\end{document}